\documentclass[tightenlines,showpacs,nofootinbib]{revtex4}
\usepackage{graphicx}
\usepackage{epsfig}
\usepackage{bm}

\makeatletter \leftline{\epsfbox{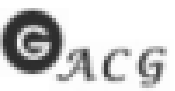}}
 \leftline
{\footnotesize{\textbf{\textsf{GACG-04/13}}}}
\begin{document}
\title{Quasinormal modes, Superradiance and Area Spectrum for $2+1$ Acoustic
Black Holes}
\author{Samuel Lepe}
\email{slepe@gacg.cl}
\author{Joel Saavedra}
\email{joelium@gacg.cl}
\affiliation{Instituto de F\'{\i}sica, Pontificia Universidad Cat\'olica de Valpara\'{\i}%
so, \\
Casilla 4950, Valpara\'{\i}so, Chile.}

\begin{abstract}
We present an exact expression for the quasinormal modes of acoustic
disturbances in a rotating $2+1$ dimensional sonic black hole (draining
bathtub fluid flow) in the low frequency limit and evaluate the adiabatic
invariant proposed by Kunstatter. We also compute,via Bohr-Sommerfeld
quantization rule the equivalent area spectrum for this acoustic black hole,
and we compute the superradiance phenomena for pure spinning $2+1$ black
holes.
\end{abstract}

\maketitle
\preprint{GACG-04/13}
\affiliation{Instituto de F\'{\i}sica, Pontificia Universidad Cat\'olica de Valpara\'{\i}%
so, \\
Casilla 4950, Valpara\'{\i}so, Chile.}

\section{\label{sec:Int}Introduction}

Analog models of general relativity have received great attention in the
last few years, since it is believed that these models are shedding light on
possible experimental verifications of some fundamental problems in black
hole physics, such as the evaporation of black holes and semiclassical
quantities. The idea of using supersonic acoustic flows as analog systems to
mimic some properties of black hole physics was proposed for the first time
by Unruh \cite{Unruh:81} (for a comprehensive review of Analog Black Holes
see \cite{Novello:2002qg} and references therein, and for a pedagogical
review see \cite{Visser:1998qn}\cite{Visser:1997ux}\cite{Visser:1993ub}).
The basis of the analogy between gravitational black hole and sonic black
holes comes from considering the propagation of acoustic disturbances on a
barotropic, inviscid, inhomogeneous and irrotational (at least locally)
fluid flow. \ It is well known that the equation of motion for this acoustic
disturbance (described by its velocity potential $\psi $) is identical to
the Klein-Gordon equation for a massless scalar field minimally coupled to
gravity in a curved spacetime \cite{Visser:1998qn}\cite{Visser:1997ux}\cite%
{Visser:1993ub}.

Therefore, this analogy gives us a powerful terrestrial laboratory to
explore principal aspects of black hole physics, for example, the Hawking
radiation with its significant consequence about the paradox of information
loss, quasinormal modes (QNMs), known as the ''ringing'' of black holes,
also play an important role in classical aspects of black holes physics. In
particular, from Hod's proposal \cite{Hod:1998vk}, modes with high damping
have received great attention, specially their relevance in the quantization
of the area of black holes and the possibilities of fixing the Immirzi $%
\gamma $ parameter in loop quantum gravity \cite{Dreyer:2002vy} \cite%
{Kunstatter:2002pj}. In principle all these processes can be studied in a
laboratory using the analog models of gravity.

In this work we analytically compute the QNMs or acoustic disturbances in a
rotating $2+1$ dimensional sonic black hole (draining bathtub fluid flow),
following the methodology developed in Refs.\cite{Basak:2003uj} and \cite%
{Basak:2002aw}. We also compute the adiabatic invariant proposed by
Kunstatter \cite{Kunstatter:2002pj}, and we compute the equivalent area
spectrum for this acoustic black hole via Bohr-Sommerfeld quantization rule.
For a $2+1$ dimensional pure spinning black hole called perfect vortex \cite%
{Fischer:2001jz} we show explicitly the absence of superresonance phenomena
and the absence of QNMs. We conjecture that this acoustic space behaves like
the BTZ extremal black hole \cite{Crisostomo:2004hj}.

The organization of the paper is as follows: In Sec. II we specify the
Draining Vortex, the notation and some useful quantities for this acoustic
geometry. In Sec. III we determine the QNMs, mass and the area spectrum in
the large damping limits. In Sec. IV we treat the perfect vortex and
explicty compute the absence of the superresonance and QNMs for this
acoustic black hole. Finally, we conclude in Sec. V.

\section{\label{sec:Sonicbh}Draining Vortex}

The rotating $2+1$ acoustic black hole was proposed by Visser for the first
time in Ref.\cite{Visser:1997ux} which corresponds to a draining bathtub
fluid flow with a sink at the origin. The velocity profile for this
configuration is given by
\begin{equation}
\overrightarrow{v}_{0}=-\frac{A}{r}\hat{r}+\frac{B}{r}\hat{\phi},
\end{equation}
where $A$ and $B$ are arbitrary real positive constants. If we assume the
fluid to be locally irrotational, the background velocity potential becomes

\begin{equation}
\psi _{0}=-A\log (\frac{r}{a})+B\phi ,
\end{equation}
where $a$ is some length scale (we adopt $a=1$).

The acoustic line element is given by
\begin{equation}
ds^{2}=-c^{2}dt^{2}+\left( \frac{A}{r}dt+dr\right) ^{2}+r^{2}\left( d\phi -%
\frac{B}{r^{2}}dt\right) ^{2},  \label{metric1}
\end{equation}%
being $c$ the constant velocity of sound. In order to study the
general properties of this acoustic spacetime, we perform the
coordinate transformations discussed in Ref.\cite{Basak:2002aw},
namely

\begin{eqnarray}
dt &\rightarrow &dt+\frac{1}{c^{2}}\frac{Ar}{r^{2}-\left( A/c\right) ^{2}}dr,
\\
d\phi &\rightarrow &d\phi +\frac{1}{c^{2}}\frac{AB}{r^{2}-\left( A/c\right)
^{2}}\frac{dr}{r},
\end{eqnarray}
so that the metric adopts a Kerr-like form

\begin{equation}
ds^{2}=-c^{2}\left( 1-\frac{(A/c)^{2}}{r^{2}}\right) dt^{2}+\left( 1-\frac{%
(A/c)^{2}}{r^{2}}\right) ^{-1}dr^{2}+r^{2}\left( d\phi -\frac{B}{r^{2}}%
dt\right) ^{2}.
\end{equation}%
The radii of \ the horizon and ergosphere for this acoustic black hole are
respectively
\begin{eqnarray}
r_{+} &=&\frac{A}{c}, \\
r_{e} &=&\sqrt{r_{+}^{2}+\left( B/c\right) ^{2}}.
\end{eqnarray}%
By considering the Komar \cite{Basak:2003yt} integrals we can
calculate two \textquotedblright conserved
charges\textquotedblright , conserved in the sense that they are
generated by divergenceless currents
\begin{equation}
M=\frac{c^{2}}{2}\left[ 1+\left( B/A\right) ^{2}\right] ,  \label{mass1}
\end{equation}

\begin{equation}
J=\frac{1}{4}c^{2}B.  \label{angular1}
\end{equation}%
Though we have labelled these conserved charges $M$ and $J$
 they do not represent the physical mass and angular
momentum of the spacetime, such an identification would require
the use of the Einstein equations, which do not apply in the
current context. It is straightforward to take the limit of static
acoustic black holes ($B=0$) for these conserved charges. But if
we take the limit of pure spinning acoustic black holes (perfect
vortex $A=0$ in (\ref{metric1})) the $M$ charge is undefined and
we must use an independent method such an used for extremal black
holes in general relativity.

Other interesting quantities for the study of black hole physics are the
angular velocity of the horizon

\begin{eqnarray}
\Omega _{+} &=&\frac{4}{c^{2}}Jr_{+}^{-2} \\
&=&\frac{c^{4}}{4J}\left( \frac{2M}{c^{2}}-1\right) ,
\end{eqnarray}
and the surface gravity

\begin{equation}
K_{+}=c^{2}r_{+}^{-1},
\end{equation}
which allows to define the analogous to the Hawking temperature:

\begin{equation}
T_{H}=\frac{1}{2\pi c}K_{+}.
\end{equation}%
In the quantum version of this system we can hope that the acoustic black
hole emits \textquotedblright acoustic Hawking radiation\textquotedblright .
This effect coming from the horizon of events is a pure kinematical effect
that occurs in any Lorenzian geometry independent of its dynamical content%
\cite{Visser:1997ux}. It is well known that the acoustic metric does not
satisfy the Einstein equations, due to the fact that the background fluid
motion is governed by the continuity and the Euler equations. As a
consequence of this fact, one should expect that the thermodynamic
description of the acoustic black hole is ill defined. However, this
powerful analogy between black hole physics and acoustic geometry allows to
extend the study of many physical quantities associated to black holes, such
as quasinormal modes and area spectrum. In the next section we consider this
aspect.

\section{\label{sec:Sonicbh1}QNMs and Area Spectrum of Draining Vortex}

The basis of the analogy between Einstein black holes and sonic black holes
comes from considering the propagation of acoustic disturbances on a
barotropic, inviscid, inhomogeneous and irrotational fluid flow. It is well
known that the equation of motion for these acoustic disturbances (described
by its velocity potential $\psi $) is identical to the Klein-Gordon equation
for a massless scalar field minimally coupled to gravity in a curved
spacetime \cite{Visser:1998qn}\cite{Visser:1997ux}\cite{Visser:1993ub},
\textit{i.e.}

\begin{equation}
\frac{1}{\sqrt{g}}\partial _{\mu }\left( \sqrt{g}g^{\mu \nu }\partial _{\nu
}\Phi \right) =0.
\end{equation}
We make the Ansatz
\begin{equation}
\Psi (t,r,\phi )=R(r)~e^{-i\omega t}~e^{im\phi },  \label{psi}
\end{equation}
where $m$ is a real constant and in order to make $\Psi (t,r,\phi )$ single
valued, $m$ must take on integer values.\newline
The radial function $R(r)$ satisfies \cite{Basak:2002aw} the following
equation:
\begin{equation}
\frac{d^{2}R(r)}{d{r}^{2}}+P_{1}(r)~\frac{dR(r)}{dr}+Q_{1}(r)~R(r)=0~,
\label{Rr}
\end{equation}
where
\begin{equation}
P_{1}\left( r\right) =\frac{1}{r(r^{2}-r_{+}^{2})}\left( r^{2}+r_{+}^{2}+2i%
\frac{r_{+}}{c}\left( \Omega _{+}r_{+}^{2}m-\omega r^{2}\right) \right) ,
\label{p1}
\end{equation}
and
\begin{eqnarray}
Q_{1}\left( r\right) &=&\frac{1}{r^{2}(r^{2}-r_{+}^{2})}\left( 2i\frac{%
\Omega _{+}}{c}r_{+}^{3}m+\left( r^{2}-\left( \frac{\Omega _{+}}{c}\right)
^{2}r_{+}^{4}\right) m^{2}\right. +  \nonumber \\
&&+\left. 2\frac{\Omega _{+}}{c}\frac{\omega }{c}r_{+}^{2}r^{2}m-\left(
\frac{\omega }{c}\right) ^{2}r^{4}\right) .  \label{q}
\end{eqnarray}

Solutions of (\ref{Rr}) were found in Ref.\cite{Basak:2002aw} using an
adaptation of the matching procedure developed by Starobinsky. In order to
solve the radial equation (\ref{Rr}) they considered a new radial function
defined by

\begin{equation}
R(r)=\frac{r}{r_{+}}\exp \left[ \frac{i}{2}\frac{r_{+}}{c}\left( \omega \log
\left( \frac{r^{2}}{r_{+}^{2}}-1\right) -m\Omega _{+}\log \left( 1-\frac{%
r_{+}^{2}}{r^{2}}\right) \right) \right] ~~L(r)~.
\end{equation}%
and then, through a matching procedure for the region $\omega \left(
r^{2}/r_{+}^{2}-1\right) \ll m2\pi T_{H}$ and $\omega \ll 2\pi T_{H}$, the
radial equation for $L(r)$ is transformed in the Riemann-Papparitz equation
whose solutions in terms of hypergeometric functions read as follows

\begin{eqnarray}
L(r) &=&\left( \frac{r^{2}}{r_{+}^{2}}-1\right) ^{\alpha ^{\prime }}\left(
\frac{r}{r_{+}}\right) ^{2\beta ^{\prime }}~\left\{ C_{1}~_{2}F_{1}\left(
\alpha ;\beta ;\gamma ;1-\frac{r^{2}}{r_{+}^{2}}\right) \right.  \nonumber \\
&&-\left. C_{2}\left( \frac{r^{2}}{r_{+}^{2}}-1\right) ^{1-\gamma
}~_{2}F_{1}\left( \alpha +1-\gamma ;\beta +1-\gamma ;2-\gamma ;\frac{r^{2}}{%
r_{+}^{2}}-1\right) \right\} ,
\end{eqnarray}
where

\begin{eqnarray}
\alpha ^{^{\prime }} &=&-~i~\frac{Q}{2}~~,~~\beta ^{^{\prime }}=-\frac{1}{2},
\nonumber \\
&& \\
\alpha &=&-\frac{S+~iQ}{2}~~,~~\beta =\frac{S-~i~Q}{2}~~,~~\gamma =1-i~Q,
\nonumber
\end{eqnarray}
and
\begin{eqnarray}
S^{2} &=&m^{2}-2\left( \frac{1}{2\pi T_{H}}\right) ^{2}\omega \left( \omega
-m\Omega _{+}\right) , \\
Q^{2} &=&\left( \frac{1}{2\pi T_{H}}\right) ^{2}\left( \omega -m\Omega
_{+}\right) ^{2}.
\end{eqnarray}
\bigskip

In order to compute the QNMs, we need to impose the boundary conditions upon
the solution of the radial equation, meaning that only purely ingoing waves
are allows at the horizon, that is, $C_{2}=0$ (since nothing comes out of
the horizon) while $L(r)$ is given by
\begin{equation}
L\left( r\right) =C_{1}\left( \frac{r^{2}}{r_{+}^{2}}-1\right) ^{\alpha
^{\prime }}\left( \frac{r}{r_{+}}\right) ^{2\beta ^{\prime
}}~~_{2}F_{1}\left( \alpha ;\beta ;\gamma ;1-\frac{r^{2}}{r_{+}^{2}}\right) .
\end{equation}%
We also need to know the behavior of $L(r)$ in the asymptotic region ($%
r\rightarrow \infty $)

\[
L\left( r\right) =D_{1}(\frac{r^{2}}{r_{+}^{2}}-1)^{\left( S-1\right)
/2}+D_{2}(\frac{r^{2}}{r_{+}^{2}}-1)^{-\left( S+1\right) /2},
\]
where the constant coefficients $D_{1}$ and $D_{2}$ are given by

\begin{eqnarray}
D_{1} &=&\frac{\Gamma \left( \gamma \right) \Gamma \left( \beta -\alpha
\right) }{\Gamma \left( \beta \right) \Gamma \left( \gamma -\alpha \right) }%
C_{1}, \\
D_{2} &=&\frac{\Gamma \left( \gamma \right) \Gamma \left( \alpha -\beta
\right) }{\Gamma \left( \beta \right) \Gamma \left( \gamma -\beta \right) }%
C_{1}.
\end{eqnarray}
We require that only outgoing waves should be present at the infinity. To
see this condition we consider the possible values for $S$:

\begin{eqnarray}
S &>&1\Longrightarrow D_{1}=0\text{ }\left( \beta =-n,\text{ }\gamma -\alpha
=-n\right) , \\
S &<&-1\Longrightarrow D_{2}=0\text{ }\left( \alpha =-n,\text{ }\gamma
-\beta =-n\right) .
\end{eqnarray}
This set of conditions can be summarized in the following condition:

\begin{equation}
\frac{1}{2}\left( S-iQ\right) =-n,  \label{ndef}
\end{equation}
where $n$ is a positive integer. Then, from (\ref{ndef}) we obtain the
frequencies of the QNMs

\begin{equation}
\alpha =i2n+\sqrt{\left( m\beta -i2n\right) ^{2}+m^{2}},
\end{equation}
being $\alpha =\omega /2\pi T_{H}$ and $\beta =\Omega _{+}/2\pi T_{H}$.

In particular, the case of highly damped modes (\textit{i.e.} QNMs with a
large imaginary part) has received great attention in the last time
specially regarding their relevance in the quantization of the area of black
holes and the possibilities of fixing the Immirzi \ $\gamma $ parameter in
loop quantum gravity through QNMs. In the present case and as a consequence
of the Starobinsky matching procedure meaning $\omega /2\pi T_{H}\ll 1$,
i.e., $\alpha \ll 1$, the spectrum of frequencies for $n\gg 1$ becomes
\begin{equation}
\omega =m\Omega _{+},
\end{equation}
and the imaginary part of the spectrum is vanishes, a result that differs
from the gravitational case. Let us note that in order to satisfy $\omega
/2\pi T_{H}\ll 1$, we also must have $\beta \ll 1\leftrightarrow \Omega
_{+}\ll 2\pi T_{H}/m$.

On the other hand, in the sense of Hod \cite{Hod:1998vk}, \cite{Hod:2003jn}
the real part of the spectrum of QNMs of the black holes does not give the
correct Schwarzschild limit $4T_{H}ln3+m\Omega _{+}$. The absence of the
Schwarzschild asymptotic limit was claimed in Ref.\cite{Berti:2004ju} and
its absence can be understood because the Hod's conjecture is fundamentally
based on black hole thermodynamics. At this point, we remark that the sonic
metric does not satisfy the Einstein equations. Therefore, a thermodynamics
scheme of acoustic black holes inspired from black hole physics in general
relativity does not apply in the acoustic case.

Now, we consider Re$\left( \omega \right) $ to be a fundamental vibrational
frequency for a black hole of energy $E=M$. Given a system with energy $E$
and vibrational frequency $\omega $ one can show that the quantity

\begin{equation}
I=\int {\frac{dE}{\omega (E)}},  \label{bohr}
\end{equation}
is an adiabatic invariant \cite{Kunstatter:2002pj} which via Bohr-Sommerfeld
quantization, has an equally spaced spectrum in the semi-classical limit:

\begin{equation}
I \approx k\hbar.  \label{smi}
\end{equation}

Now by taking Re$\left( \omega \right)\equiv \omega _{QNM}$ in this context
we have ($J$ fixed)

\begin{eqnarray}
I &=&\int \frac{dM}{\omega _{QNM}}, \\
&=&\frac{2J}{c^{2}}\frac{1}{m}\ln \left( \frac{2M}{c^{2}}-1\right) ,
\end{eqnarray}%
and the mass spectrum becomes

\begin{equation}
M\left( m,k\right) =\frac{c^{2}}{2}\left[ 1+\exp \left( \frac{c^{2}}{2J}%
mk\hbar \right) \right] .
\end{equation}
On the other hand, the perimeter of the horizon of the black hole is given by

\begin{eqnarray}
A_{+} &=&2\pi r_{+}, \\
&=&\frac{8\pi J}{c^{3}}\left( \frac{2M}{c^{2}}-1\right) ^{-1/2},
\end{eqnarray}
and as a consequence of the mass spectrum, the area spectrum becomes
\begin{equation}
A_{+}\left( m,k\right) =\frac{8\pi }{c^{3}}J\exp \left[ -\left( \frac{c}{2}%
\right) ^{2}\frac{m}{J}\left( k\hbar \right) \right] ,
\end{equation}
where we can see that both mass spectrum and area spectrum are not equally
spaced against to the Schwarzschild black hole \cite{Padmanabhan:2003fx}
\cite{Choudhury:2003wd}. The conserved charges $M$ and $J$ are not
independent in the acoustic case (\ref{mass1}, \ref{angular1}). Since the
constant $B$ establishes their relationship. If we express these charges in
terms of $\Omega _{+}$ and $T_{H}$ (physical parameters of sonic black hole)

\begin{eqnarray}
M &=&\frac{c^{2}}{2}\left[ 1+\left( \frac{\Omega _{+}}{2\pi T_{H}}\right)
^{2}\right] , \\
J &=&\frac{1}{4}c^{2}A\left( \frac{\Omega _{+}}{2\pi T_{H}}\right) ,
\end{eqnarray}%
then the parameter $A$ which is an arbitrary constant, can be freely defined
and hence as a consequence of this we are left with a unique effective
degree of freedom ($M$) which allows an unambiguous computation of the
adiabatic invariant. Let us recall however that the present situation is not
applicable to the BTZ \cite{mann} \cite{Setare:2003hm} of Kerr black holes
\cite{Setare:2003bd}\cite{Das:2004ak}\cite{Setare:2004uu} . In these later
ones, $M$ and $J$ are effectively independent charges with no relation
between them as described above.

\section{\label{sec:Sonicbh2}Perfect Vortex}

The perfect vortex is obtained when we replace $A=0$ in (\ref{metric1}). In
this case this spacetime (acoustic vortex) represents a fluid with a
non-radial flow. This geometry was studied from a Riemannian geometry point
of view in Ref. \cite{Fischer:2001jz}. The metric and its inverse are,
respectively, given by

\begin{equation}
g_{\mu \nu }=\left(
\begin{array}{ccc}
-c^{2}\left(1+ r_{e}^{2}/r^{2}\right) & 0 & -c~r_{e} \\
0 & 1 & 0 \\
-c~r_{e} & 0 & r^{2}%
\end{array}
\right) ,  \label{perfectmetric1}
\end{equation}

\begin{equation}
g^{\mu \nu }=\left(
\begin{array}{ccc}
-1/c^{2} & 0 & -r_{e}/c~r^{2} \\
0 & 1 & 0 \\
-r_{e}/c~r^{2} & 0 & \left( 1-r_{e}^{2}/r^{2}\right) /c~r^{2}%
\end{array}
\right) ,  \label{perfectmetric2}
\end{equation}
where
\begin{equation}
r_{e}=\frac{B}{c},
\end{equation}
is the radius of the ergosphere and $c$is the constant velocity of
sound as referenced in Refs. \cite{Basak:2003uj}
\cite{Basak:2002aw}.

Therefore using the Klein-Gordon equation with (\ref{perfectmetric1}) and
considering solutions of the form,
\begin{equation}
\Psi (t,r,\phi )=R(r)~e^{-i~\omega ~t}~e^{i~m~\phi },  \label{psi1}
\end{equation}
where$~m~$is an integer constant, we obtain the radial equation

\begin{equation}
\frac{d^{2}R\left( r\right) }{dr^{2}}+\frac{1}{r}\frac{dR\left( r\right) }{dr%
}+\frac{1}{r^{2}}\left[ \left( \frac{r_{e}~m}{r}\right) ^{2}-\left(
m^{2}+2r_{e}\frac{\omega }{c}~m\right) +\left( \frac{\omega ~r}{c}\right)
^{2}\right] R\left( r\right) =0.  \label{Rperfect}
\end{equation}
With a change of variable

\begin{equation}
R\left( r\right) =r^{-1/2}H\left( r\right) ,
\end{equation}
the radial equation (\ref{Rperfect}) can be written as
\begin{equation}
\frac{d^{2}H\left( r\right) }{dr^{2}}+\left[ \frac{1}{c^{2}}\left( \omega -%
\frac{c~r_{e}}{r^{2}}m\right) ^{2}-\left( m^{2}-\frac{1}{4}\right) \frac{1}{%
r^{2}}\right] H\left( r\right) =0.  \label{h(r)}
\end{equation}
First, we shall study the super-resonance (analog to the super-radiance in
black hole physics). For this goal we take the limit $r\rightarrow \infty $,
and (\ref{h(r)}) adopts the form
\begin{equation}
\frac{d^{2}H_{\infty }\left( r\right) }{dr^{2}}+\left( \frac{\omega }{c}%
\right) ^{2}H_{\infty }\left( r\right) =0,
\end{equation}
whose solution is

\begin{equation}
H_{\infty }\left( r\right) =\exp \left( -i\frac{\omega }{c}r\right) +%
\mathcal{R}\left( \omega ,m\right) \exp \left( i\frac{\omega }{c}r\right) ,
\end{equation}
where $\mathcal{R}\left( \omega ,m\right) $ is the reflection coefficient in
the sense of scattering potential.

Near to ergosphere ($r\rightarrow r_{e}$), the equation reads as follows
\begin{equation}
\frac{d^{2}H_{0}\left( r\right) }{dr^{2}}+\left( \frac{\omega }{c}\right)
^{2}\left( 1-\frac{\omega _{+}}{\omega }\right) \left( 1-\frac{\omega _{-}}{%
\omega }\right) H_{0}\left( r\right) =0,
\end{equation}
where
\begin{equation}
\omega _{\pm }=m\Omega _{e}\left[ 1\pm \sqrt{1-\left( 4m^{2}\right) ^{-1}}%
\right] ,
\end{equation}
and $\Omega _{e}$ is the angular velocity of the ergosphere
\begin{equation}
\Omega _{e}=\frac{c}{r_{e}}.
\end{equation}

In this region the solution reads as follow

\begin{equation}
H_{0}\left( r\right) =\mathcal{T}\left( \omega ,m\right) \exp \left( -i\frac{%
\omega }{c}\sqrt{\left( 1-\frac{\omega _{+}}{\omega }\right) \left( 1-\frac{%
\omega _{-}}{\omega }\right) }r\right) ,
\end{equation}%
where $\mathcal{T}\left( \omega ,m\right) $ is the transmission coefficient.
It is straightforward to check that the Wronskian of these approximate
solutions is constant, such that
\begin{equation}
1-\left\vert \mathcal{R}\left( \omega ,m\right) \right\vert ^{2}=\sqrt{%
\left( 1-\frac{\omega _{+}}{\omega }\right) \left( 1-\frac{\omega _{-}}{%
\omega }\right) }\left\vert \mathcal{T}\left( \omega ,m\right) \right\vert
^{2}.
\end{equation}%
We can observe the absence of the super-resonance for the perfect vortex,
meaning that it is impossible to have $\left\vert \mathcal{R}\left( \omega
,m\right) \right\vert ^{2}>1$, despite the fact that this geometry exhibits
an ergosphere. To the best our knowledge there is no gravitational black
hole exhibits this behavior. By virtue of the results report in Refs.\cite%
{Gamboa:2000uc}, \cite{Lepe:2003na} and \cite{Crisostomo:2004hj},
we could make the conjecture that the extreme BTZ black hole could
be a possible candidate \cite{SLJS}. At this point we notice that
our results are valid when the local velocity of the sound is a
constant. Certainly this fact is different from the physical
situation found in Bose-Einstein Condensates (BEC) where it is
well known that the sound velocity is proportional to the density
\cite{bec1} \cite{bec2}. The relevance of the perfect vortex in
the case of $c$ constant was discussed for the first time in the
Ref. \cite{Fischer:2001jz} where the acoustic perfect vortex leads
to deflection of phonons like the fotons bending due to the
gravitational field. Recently, in the case $c^{2}(r)\sim \rho$
Ref.\cite{Slatyer:2005ty}\cite{Basak:2005fv} \ it was shown that
the superresonant scattering is possible if an ad hoc density
profile is given.

In the following we explore the existence of QNMs. In order to do this, we
need to solve (\ref{Rperfect}) explicitly. Let us start with the change of
variables $x=r^{2}/r_{e}^{2}$. Equation (\ref{Rperfect}) becomes

\begin{equation}
\frac{d^{2}R\left( x\right) }{dx^{2}}+\frac{1}{x}\frac{dR\left( x\right) }{dx%
}+\frac{1}{4}\left[ \frac{m^{2}}{x}-\left( m^{2}+\frac{2\omega }{\Omega _{e}}%
m\right) +\left( \frac{\omega }{\Omega _{e}}\right) ^{2}x\right] \frac{1}{%
x^{2}}R\left( x\right) =0,  \label{RXperfect}
\end{equation}
and, in the spirit of the Starobinsky matching procedure, we consider the
region described by
\begin{equation}
\frac{\omega }{\Omega _{e}}x\ll m\text{ \ \ and \ \ }\frac{\omega }{\Omega
_{e}}\ll 1.\text{\ }  \label{wcondition}
\end{equation}
In this region (\ref{RXperfect}) reads as follows

\begin{equation}
\frac{d^{2}R\left( x\right) }{dx^{2}}+\frac{1}{x}\frac{dR\left( x\right) }{dx%
}+\frac{1}{4}\left[ \frac{m^{2}}{x}-\left( m^{2}+\frac{2\omega }{\Omega _{e}}%
m\right) \right] \frac{1}{x^{2}}R\left( x\right) =0,\label{rrrx}
\end{equation}
whose solutions are given by

\begin{equation}
R(r)=C_{1}J_{-S}\left( mr_{e}r^{-1}\right) +C_{2}Y_{-S}\left(
mr_{e}r^{-1}\right) ,
\end{equation}
where $J_{-S}$, $Y_{-S}$ are Bessel functions, and $S$ is given by

\[
S=m\sqrt{1+\frac{2}{m}\frac{\omega }{\Omega _{e}}}.
\]
>From the asymptotic behaviors \cite{abramowitz} of these Bessel functions,
it is straightforward to show that

\begin{eqnarray}
\lim_{r\rightarrow \infty }R\left( r\right) &\rightarrow &\frac{1}{\Gamma
\left( 1-S\right) }\left( \frac{2r}{mr_{e}}\right) ^{S}\left[
C_{1}-C_{2}\cot \left( S\frac{mr_{e}}{r}\right) \right] \\
&&+C_{2}\frac{1}{\Gamma \left( 1+S\right) }\frac{1}{\sin \left( S\frac{mr_{e}%
}{r}\right) }\left( \frac{mr_{e}}{2r}\right) ^{S}.  \label{besselinf}
\end{eqnarray}
Then, imposing the regularity conditions to find the QNMs, we obtain

\begin{equation}
\Gamma \left( 1-S\right) =\infty \Longrightarrow S=n=1,2,3,...
\end{equation}
Finally, we find

\begin{equation}
\omega =\left( \frac{n^{2}-m^{2}}{2m}\right) \Omega _{e},
\end{equation}%
and notice that, the imaginary part of the frequencies is
vanishing, showing the absence of QNMs for the perfect vortex.
This behavior is analogous to the case of BTZ extreme black hole
\cite{Crisostomo:2004hj}. Additionally,  there is an infinite
tower of ordinary well behaved normal modes in the limit of the
Starobinsky matching procedure and it is straightforward to show
that this spectrum contains only positive frequencies  according
to (\ref{wcondition}) (see Appendix).

\section{\label{sec:Sonicbh4}Conclusions and Remarks}

In this paper we have studied QNMs, mass spectrum and area spectrum for the $%
2+1$ acoustic black hole, called 'draining bathtub fluid flow' or vortex
geometry. We also studied superresonance phenomena and QNMs for the perfect
vortex metric.

In the draining bathtub fluid flow we have found the explicit expression of
the QNMs spectrum, and for highly damping modes this spectrum contains only
real frequencies. We note that this last result is similar to the one
proposed by Hod in Ref.\cite{Hod:2003hn} for the asymptotic real spectrum of
the QNMs of rotating black holes (Kerr black holes). However, sonic black
holes do not approach the Schwarzschild limit conjectured by Hod. \ This
absence of the Schwarzschild limit was also pointed in Ref.\cite%
{Berti:2004ju} showing that the applications of Hod's conjecture to the
acoustic black hole seems impossible. At this point we can note that this
absence is natural, because Hod's conjecture is deeper based on
thermodynamics black hole physics. Finally, we showed that the area spectrum
and mass spectrum are not equally spaced.

Along the same lines we computed the perfect vortex describes in Ref.\cite%
{Basak:2003uj}. Explicit solutions of the radial equations were obtained and
we have showed that the superresonace modes can not be excited, despite the
fact that the acoustic geometry has an ergosphere. We also demonstrated the
absence of QNMs for the perfect vortex and we conjecture that this acoustic
black hole behaves like an extremal BTZ black hole.

\begin{acknowledgments}
We are grateful to E. Ayon for many enlightening discussions. We
also acknowledge the referee for useful suggestions in
 order to improve the presentation of the results of this paper.
This work was supported by COMISION NACIONAL DE CIENCIAS Y
TECNOLOGIA through FONDECYT \ Grant 1040229 (SL) and Postdoctoral
Grant 3030025 (JS). This work was also partially supported by PUCV
Grant No. 123.779/2005 (SL) and No. 123.778/2005 (JS). The authors
wish to thank Centro de Estudios Cient\'{\i}ficos (CECS) and
Departamentos de F\'{\i}sica de la Universidad de La Frontera y de
la Universidad de Concepci\'on for its kind hospitality. We thank
U. Raff for a careful reading of the manuscript.
\end{acknowledgments}

\section{Appendix}

From the equation (\ref{rrrx})

\begin{equation}
\frac{d^{2}R\left( x\right) }{dx^{2}}+\frac{1}{x}\frac{dR\left( x\right) }{dx%
}+\frac{1}{4}\left[ \frac{m^{2}}{x}-S^{2}\left( m\right) \right] \frac{1}{%
x^{2}}R\left( x\right) =0,
\end{equation}

where

\begin{equation}
S^{2}\left( m\right) =m^{2}\left[ 1+2\frac{\omega }{\Omega _{e}}\frac{1}{m}%
\right] ,
\end{equation}%
it can directly be shown that in order to satisfy the regularity
condition for the QNMs, $S^{2}\left( m\right)$  must be positive,
such that only outgoing waves are allowed at infinity (the
asymptotic solutions are given by (\ref{besselinf})). For $m<0$\
we can write $S^{2}(m)$ in the following form

\begin{equation}
S^{2}\left( m<0\right) =\left\vert m\right\vert ^{2}\left[ 1-2\frac{\omega }{%
\Omega _{e}}\frac{1}{\left\vert m\right\vert }\right] ,
\end{equation}%
where  the condition

\begin{equation}
0<\omega <\frac{1}{2}\left\vert m\right\vert \Omega _{e},  \label{condicion1}
\end{equation}%
 is consistent with the range of validity of the Starobinsky
matching procedure given by (\ref{wcondition}).

Now we study the behavior of the infinite tower of normal modes
given by

\begin{equation}
\omega =\frac{1}{2}\left( \frac{n^{2}-m^{2}}{m}\right) \Omega _{e}.
\end{equation}

 For $m<0$ this equation can be rewritten as

\begin{equation}
\omega =-\frac{1}{2}\left( \frac{n^{2}-\left\vert m\right\vert ^{2}}{%
\left\vert m\right\vert }\right) \Omega _{e}=\frac{1}{2}\left\vert
m\right\vert \Omega _{e}-\frac{n^{2}}{2\left\vert m\right\vert }\Omega _{e},%
\text{ }
\end{equation}%
and in order to satisfy (\ref{wcondition}) it is necessary that
$\left\vert m\right\vert >n$.

The critical case $m>n$ admits the existence of negative
frequencies such that it can be associated with a possible
instability. But according to (\ref{wcondition}) this instability
does not appear. This shows the robustness of the matching
procedure developed by Starobinsky from which we can obtain a well
behaved tower of ordinary normal modes.

\end{document}